\documentclass[prl,
twocolumn,
11pt,
nofootinbib]{revtex4}

\usepackage{graphicx}
\usepackage{amsmath, amssymb, amsfonts}
\usepackage[usenames]{color}
\usepackage{color,bm}

\usepackage[a4paper,inner=2cm,outer=1.5cm,top=1.5cm,bottom=2.0cm]{geometry}
\usepackage[colorlinks=true, linkcolor=navyblue, urlcolor=navyblue, citecolor=darkred]{hyperref}
\definecolor{navyblue}{rgb}{0, 0.0, 1.0}
\definecolor{darkred}{rgb}{1.0, 0.0, 0.0}

\newcommand{\vecc}[1]{\mbox{\boldmath $#1$}}
\newcommand{\ba}{\begin{eqnarray}}   
\newcommand{\ea}{\end{eqnarray}}
\newcommand{\be}{\begin{equation}}
\newcommand{\ee}{\end{equation}}

\newcommand\nn{\nonumber}
\newcommand{\bc}{\begin{center}} 
\newcommand{\ec}{\end{center}}   

\begin{document}

\title{Angular Dependence of the Polarization Transfer \\
to the Proton  in the $e \vec p \to e \vec p$ Process }

\author{M. V. Galynskii}
\email {galynski@sosny.bas-net.by}
\affiliation{Joint Institute for Power and Nuclear Research -- Sosny,
National Academy of Sciences of Belarus, Minsk, 220109 Belarus
}

\begin{abstract}
The dependence of the longitudinal polarization transfer to the proton in the $ e \vec p \to e \vec p$
process on the proton scattering angle has been numerically analyzed for the case where the initial
(at rest) proton is partially polarized along the direction of motion of the detected recoil proton.
The analysis is based on the results of JLab's polarization experiments on measuring the ratio
of the Sachs form factors in the $\vec e  p \to e \vec p$ process, using the Kelly (2004)
and Qattan (2015) parameterizations for their ratio, in the kinematics of the
SANE Collaboration experiment (2020) on measuring the double spin asymmetry in the
$\vec e \vec p \to e p$ process. It is shown that a violation of the scaling of the Sachs
form factors leads to a significant increase in the magnitude of the polarization transfer
to the proton in comparison with the case of dipole dependence.
\end{abstract}


\maketitle

{\bf Introduction.} The proton electric ($G_{E}$) and magnetic ($G_{M}$) form factors, the so-called
Sachs form factors (SFFs), have been experimentally studied since the 1950s in the
elastic scattering of unpolarized electrons by a proton. All experimental data
on the behavior of the SFFs have been obtained using the Rosenbluth technique based
on the application of the Rosenbluth cross section (within the one-photon exchange
approximation) for the $ep\to ep$ process in the rest frame of the initial proton
\cite{Rosen}:
\ba
\label{Ros}
\frac{d\sigma} {d\Omega_e}=
\frac{\alpha^2E_2\cos^2(\theta_e/2)}{4E_1^{3}\sin^4(\theta_e/2)}
\frac{1}{1+\tau_p} \left(G_E^{2}
+\frac{\tau_p}{\varepsilon}G_M^{2}\right).
\ea
Here, $\tau_p=Q^2/4M^2, Q^2=4E_1 E_2\sin^2(\theta_e/2)$ is the squared
4-momentum transfer to the proton; $M$ is the proton mass; $E_1$ and $E_2$
are the energies of the initial and final electrons, respectively;
$\theta_e$ is the electron scattering angle; $\varepsilon$ is the degree of linear
(transverse) polarization of the virtual photon \cite{Dombey,Rekalo74,AR,GL97};
$\alpha=1/137$ is the fine structure constant.

As follows from Eq. (\ref{Ros}), the main contribution to the cross section
for the $ep \to ep$ process at large $Q^2$ values is from the term proportional to
$G_M^{\,2}$; hence, the extraction of the contribution of $G_E^{\,2}$ even at
$Q^2\geqslant 2$ GeV$^2$ is a challenging problem \cite{ETG15,Punjabi2015}.

Using the Rosenbluth technique, the dipole dependence of the SFF
on $Q^2$ in the range $Q^2\leqslant6$ GeV$^2$ was established \cite{ETG15,Punjabi2015}.
The $G_E$ and $G_M$ values turned out to be connected by the scaling relation
$G_M\approx \mu_p G_E$, as a result of which the following approximate equality is
valid for the ratio $R \equiv \mu_p G_E/G_M$ ($\mu_p=2.79$ is the proton
magnetic moment):
\be
R \approx 1.
\label{Rffs}
\ee

Akhiezer and Rekalo \cite{Rekalo74} proposed a method for measuring the $R$ ratio
based on the phenomenon of polarization transfer from the initial electron to the
final proton in the $\vec e  p \to e \vec p$ process. The precision experiments based
on this method, performed at the Thomas Jefferson National Accelerator Facility
(JLab, United States) \cite{Jones00,Gay01,Gay02}, revealed that the $R$ ratio
decreases rapidly with an increase in $Q^2$, which indicates violation of the SFF
dipole dependence (scaling). This decrease turned out to be linear in the range
$0.4 \; \rm {GeV^2} \leqslant Q^2 \, \leqslant 5.6$ GeV$^2$.

Measurements of the $R$ ratio, repeated with a higher accuracy
\cite{Pun05,Puckett10,Puckett12,Puckett17,Qattan2005}
in a wide range of $Q^2$ values up to 8.5 GeV$^2$, using both the Akhiezer--Rekalo
method \cite{Rekalo74} and the Rosenbluth technique \cite{Qattan2005}, only
confirmed the discrepancy of the results.

Experimental values of $R$ were obtained by the SANE collaboration \cite{Liyanage2020}
using the third approach \cite{Donnelly1986}: they were extracted from the measurements of
the double spin asymmetry in the $\vec e \vec p \to e p$ process  in the case of partial
polarization of the electron beam and proton target. The degree of polarization of the
proton target ($P_t$ ) was $P_t=(70 \pm 5)$ \%. The experiment was performed at two electron
beam energies $E_1$, 4.725 and 5.895 GeV, and two $Q^2$ values, 2.06 and 5.66 GeV$^2$.
The $R$ values extracted in \cite{Liyanage2020} are in agreement with the results of the previous
polarization experiments carried out at the JLab
\cite{Jones00,Gay01,Gay02,Pun05,Puckett10,Puckett12,Puckett17}.

The fourth method for measuring the $R$ ratio was proposed in
\cite{JETPL2008,JETPL18,JETPL19,JETPL2021,PEPAN2022}; it is based
on the polarization transfer from the initial to the final proton
in the elastic process
\ba
e(p_1)+ p(q_1,s_1) \to e(p_2)+p(q_2,s_2)
\label{EPEP}
\ea
in the case where initial (at rest) and final protons are fully or partially polarized
and have a common spin quantization axis, coinciding with the direction of
motion of the detected recoil proton. This method can also be applied in the
two-photon exchange approximation; it allows one to measure the squared
moduli of generalized SFFs \cite{JETPL19}. This line of research was
started in \cite{JETPL2008}.

Note that Akhiezer and Rekalo (see \cite{AR}, pp. 211--215) performed also general
calculation of the $e \vec p \to e \vec p$ cross section in the Breit system for the case of partially
polarized initial and final protons. However, they analyzed this cross section in \cite{AR}
by analogy with \cite{Rekalo74} and overlooked a more interesting case, which is considered
here and was discussed in \cite{JETPL2008,JETPL18,JETPL19,JETPL2021,PEPAN2022}.

Based on the results of JLab's polarization experiments, aimed at measuring the $R$ ratio in the
$\vec e  p \to e \vec p$ process, a numerical analysis of the $Q^2$ dependence of the ratio
of the cross sections without and with proton spin flip, the polarization asymmetry in
the $e \vec p \to e \vec p$ process, and the longitudinal polarization transfer to
the proton in the kinematics of the experiment \cite{Liyanage2020} in the case where
the initial (at rest) and final protons are polarized and have a common spin quantization axis,
coinciding with the direction of motion of the detected
final recoil proton, was performed in \cite{JETPL2021,PEPAN2022}. It was shown that the polarization
transfer to the proton is fairly sensitive to the form of the dependence of the SFF
ratio $R$  on $Q^2$, i.e., to the choice  of the parameterization for the $R$ ratio.

In this study, based on the results of JLab's polarization experiments on measuring
the SFF ratio in the $\vec e  p \to e \vec p$ process, using the Kelly \cite{Kelly2004}
(2004) and Qattan \cite{Qattan2015} (2015) parameterizations for their ratio, in the
kinematics of SANE's experiment \cite{Liyanage2020} on measuring the double spin asymmetry
in the $\vec e \vec p \to e p$ process, a numerical analysis of the dependence of the longitudinal
polarization transfer to the proton in the $e \vec p \to e \vec p$ process on the proton
scattering angle is performed for the case where the initial (at rest) proton is partially
polarized along the direction of motion of the detected recoil proton.

{\bf Cross section for the $e \vec p \to e \vec p$ process in the rest frame of the initial proton.}
Let us consider the spin 4-vectors $s_{1}$ and $s_{2}$ of the initial and final protons
with 4-momenta $q_{1}$ and $q_{2}$ in process (\ref{EPEP}) in an arbitrary reference frame.
The conditions of orthogonality ($s_{i} q_{i} = 0$) and normalization ($s_{i} ^{2} = - 1$)
make it possible to determine unambiguously the expressions for their time and space components
$s_i=(s_{i0}, \vecc s_i)$ in terms of their 4-velocities $v_i=q_i/M$ ($i=1, 2$):
\be
s_i=(s_{i0}, \vecc s_i), \; s_{i0}=\vecc v_i\, \vecc c_i, \;
\vecc s_i =\vecc c_i + \frac{(\vecc c_i \vecc v_i)\,\vecc v_i}{1+v_{i0}}\;,
\label{spinv}
\ee
where the unit 3-vectors $\vecc c_i$ ($\vecc c_i^{2}=1$) are spin quantization axes.

In the laboratory reference frame, where $q_1=(M,\vecc 0)$ and $q_2=(q_{20}, \vecc q_2)$,
the spin quantization axes $\vecc c_{1}$ and $\vecc c_{2}$ are chosen so as to coincide
with the direction of motion of the final proton:
\ba
\vecc c = \vecc c_{1} =\vecc c_{2}=\vecc n_2=  \vecc {q_2}/|\vecc q_2|\,.
\label{LSO}
\ea
Then the spin 4-vectors of the initial ($s_{1}$) and final ($s_{2}$) protons take the form
\ba
\label{DSB_LSO1}
s_1=(0,\vecc n_2 )\,, \; s_2= (|\vecc v_2|, v_{20}\, \vecc {n_2})\,,
\,\vecc n_2=  \vecc {q_2}/|\vecc q_2|\,.
\ea
The method proposed in \cite{JETPL2008,JETPL18,JETPL19,JETPL2021,PEPAN2022} is based on the
expression for the differential cross section for process (\ref{EPEP}) in the laboratory
reference frame for the case where the initial and final protons are polarized and
have a common spin quantization axis $\vecc c$ (\ref{LSO}):
\ba
\label{RosPol}
\frac{d\sigma_{\delta_1, \delta_2}} {d\Omega_e}&=&
\omega_{+} \sigma^{\uparrow\uparrow}+\omega_{-}\sigma^{\downarrow\uparrow}\,,\\
\label{RosPol2}
\sigma^{\uparrow\uparrow}&=&\sigma_M \, G^2_E ,\;\;
\sigma^{\downarrow\uparrow}=\sigma_M \frac{\tau_p}{\varepsilon} \, G^2_M\,,\\
\sigma_M&=& \frac{\alpha^2E_2\cos^2(\theta_e/2)}
{4E_1^{\,3}\sin^4(\theta_e/2)} \frac{1}{1+\tau_p}\,.
\ea
Here, $\omega_{\pm}$  are polarization factors:
\ba
\omega_{+}=(1 + \delta_1 \delta_2)/2, \, \,\omega_{-}=(1 -\delta_1 \delta_2)/2\,,
\label{omegi}
\ea
where $\delta_{1,2}$ are the doubled projections of the spins of
the initial and final protons on the spin quantization
axis $\vecc c$ (\ref{LSO}). Note that Eq. (\ref{RosPol}) is valid at
$-1\leqslant \delta_{1,2}\leqslant 1$.

Formula (\ref{RosPol}), as well as (\ref{Ros}), is the sum of two terms,
one of which contains only $G_E^{\,2}$, while the other contains
only $G_M^{\,2}$. Averaging and summing Eq. (\ref{RosPol}) over
the polarizations of the initial and final protons yields
another representation \cite{JETPL18,JETPL19} for the Rosenbluth
cross section (\ref{Ros}), $\sigma_R=d\sigma / d\Omega_e$, in the form
\ba
\label{Ross}
\sigma_R =\sigma^{\uparrow\uparrow} + \sigma^{\downarrow\uparrow}\,.
\ea
Therefore, the physical meaning of the representation of Rosenbluth formula (\ref{Ros})
as the sum of two terms, one containing only $G_E^{~2}$ and the other containing only $G_M^{~2}$,
is that it is the sum of the cross sections without ($\sigma^{\uparrow\uparrow}$)
and with ($\sigma^{\downarrow\uparrow}$) proton spin flip in the case where
the initial (at rest) proton is fully polarized along the direction of motion
of the final proton. It is often stated in the literature, including textbooks on physics
of elementary particles, that SFFs are used simply for convenience, because they make
the Rosenbluth formula simple and concise. Since these formal conclusions
about the advantages of using the SFFs can be found, in particular, in well-known
monographs published many years ago \cite{AB,BLP}, they are considered as
undoubtful and still reproduced in the literature (see, e.g., \cite{Paket2015}).

Cross section (\ref{RosPol}) can be presented in the form
\ba
&& d\sigma_{\delta_1, \delta_2}/ d\Omega_e
\label{sigma_d1_d2}
=(1+\delta_2 \delta_f) (\sigma^{\uparrow\uparrow}+\sigma^{\downarrow\uparrow}),\\
\label{delta_f}
&& \delta_f=\delta_1 (R_{\sigma}-1)/(R_{\sigma}+1), \\
&& R_{\sigma}=\sigma^{\uparrow\uparrow}/\sigma^{\downarrow\uparrow},
\label{R_sig}
\ea
where $\delta_f$ is the degree of the longitudinal polarization
of the final proton. If the initial proton is fully polarized
($\delta_1=1$), $\delta_f$ coincides with the conventional definition
of the polarization asymmetry:
\ba
\label{Asim}
A= (R_{\sigma}-1)/(R_{\sigma}+1)\;.
\ea

As follows from (\ref{RosPol2}), the ratio of the cross sections
without and with proton spin flip, $R_{\sigma}$ (\ref{R_sig}), can be
expressed in terms of the experimentally measurable
value $R \equiv \mu_p\, G_E/G_M$:
\ba
R_{\sigma}=\frac{\sigma^{\uparrow\uparrow}}{\sigma^{\downarrow\uparrow}}
=\frac{\varepsilon}{\tau_p}\,\frac{G_E^2}{G_M^2}=
\frac{\varepsilon}{\tau_p\, \mu_p^2}\, R^2.
\label{rat2}
\ea

To use the standard notation, Eq. (\ref{delta_f}) for the degree of longitudinal
polarization of the final proton can be rewritten as follows:
\ba
P_{r}= P_t (R_{\sigma}-1)/(R_{\sigma}+1),
\label{Asim_pt}
\ea
where $\delta_f$ and $\delta_1$ are replaced with $P_r$ and $P_t$, respectively.

Formula (\ref{Asim_pt}) makes it possible to express the $R$
ratio in terms of $P_r/P_t$. Indeed, having inverted the
relation in (\ref{Asim_pt}), we arrive at
\ba
R^2=\mu_p^2\,\frac{\tau_p}{\varepsilon}\, \frac{1+R_p}{1-R_p},\;
R_p=\frac{P_r}{P_t}.
\label{Rp}
\ea
The obtained Eq. (\ref{Rp}) allows one to derive $R^2$ from the results of experiments
on measuring the polarization transfer to the proton, $P_r$, in the $e \vec p \to e \vec p$
process when the initial (at rest) proton is partially polarized along
the direction of motion of the detected recoil proton.

The results of the numerical calculations of the polarization transfer $P_r$ to the proton
(\ref{Asim_pt}) as a function of the proton scattering angle both with preserved
scaling (i.e., in the case of the dipole dependence ($R=R_d$ )) and with scaling violated
are presented below; two parameterizations ($R=R_j$ and $R=R_k$) are considered:
\ba
\label{Rd}
&&R_d=1 ,    \\
&&R_j^{-1} = 1+0.1430 Q^2-0.0086 Q^4+0.0072Q^6,~~~~~~
\label{Rdj}
\ea
where the expression for $R_{j}$ (\ref{Rdj}) was proposed in \cite{Qattan2015};
$R_k$ corresponds to the Kelly parameterization \cite{Kelly2004}; the
corresponding formulas are omitted.

{\bf Kinematics of the process.}
Let us consider the dependences of the energies of the final electrons and protons on the energy of the
initial electron beam and the electron and proton scattering angles in the laboratory reference frame,
where $q_1=(M, \vecc 0)$. Using the 4-momentum conservation law $p_1+q_1=p_2 + q_2$, we obtain
expressions for the scattered electron energy $E_2$ and the squared 4-momentum
transfer to the proton $Q^2=-(q_1-q_2)^2$ as functions of the electron scattering angle $\theta_e$:
\ba
\label{E2te}
&&E_2=E_1/(1+ \frac{E_1}{M}\, (1-\cos(\theta_e)),\\
\label{Q^2te}
&&Q^2=2 E_1^{\,2}(1-\cos(\theta_e))/(1+\frac{E_1}{M}(1-\cos(\theta_e)),~~~~~
\ea
where the $\theta_e$ is the angle between the vectors $\vecc p_1$ and $\vecc p_2$,
$\cos(\theta_e)=\vecc p_1 \vecc p_2/|\vecc p_1| |\vecc p_2|$.

The final electron energy $E_2$ and proton energy $E_{2p}$
are related to $Q^2$ in the laboratory reference frame:
\ba
\label{E2Q}
&&E_2=E_1-Q^2/2M, E_{2p}=M+Q^2/2M,   \\
\label{E2tau}
&&E_2=E_1-2M\tau_p, E_{2p}=M(1+2\tau_p). 
\ea

The dependence of $E_{2p}$ and $Q^2$ on the proton scattering angle $\theta_p$, which is
the angle between the vectors $\vecc p_1$ and $\vecc q_2$,
$\cos(\theta_p)=\vecc p_1 \vecc q_2/|\vecc p_1| |\vecc q_2|$, has the form
\ba
\label{E2tp}
&&E_{2p}=M\, \frac{(E_1+M)^2+E_1^{\,2}\cos^2(\theta_p)}
{(E_1+M)^2-E_1^{\,2}\cos^2(\theta_p)}\,,\\
\label{Q^2tp}
&&Q^2=\frac{4M^2 E_1^{\,2}\cos^2(\theta_p)}
{(E_1+M)^2-E_1^{\,2}\cos^2(\theta_p)}\,.
\ea

The inverse relations between $\cos(\theta_e)$, $\cos(\theta_p)$
and $E_2$, $E_{2p}$ can be written as
\ba
\label{tetae}
&&\cos(\theta_e)=1-\frac{Q^2}{2 E_1 E_2}=1-\frac{MQ^2} {E_1(2ME_1-Q^2)},~~\\
\label{tetap}
&&\cos(\theta_p)=\frac{E_1+M}{E_1} \sqrt{\frac{\tau_p}{1+\tau_p}}.
\ea

In elastic process (\ref{EPEP}) the electron scattering angle $\theta_e$
changes from $0^{\circ}$ to $180^{\circ}$, while $Q^2$ changes in the range
of $0 \leqslant Q^2 \, \leqslant Q^2_{max}$ ($0 \leqslant \tau_p \, \leqslant \tau_{max}$), where
\ba
\label{Qmax}
Q^2_{max}=\frac{4ME_1^{\,2}}{(M+2E_1)}, \,\,
\tau_{max}=\frac{E_1^{\,2}}{M(M+2E_1)}.
\ea
Let us write the following useful relation:
\ba
\label{taum+1}
\sqrt{\frac{\tau_{max}}{1+\tau_{max}}}=\frac{E_1}{M+E_1}.
\ea

According to Eq. (\ref{Q^2te}), at $\theta_e=0$ we have $Q^2=0$
and $\tau_p=0$. However, if follows from Eq. (\ref{tetap}) that $\cos(\theta_p)=0$
in this case, which corresponds to proton scattering by $90^{\circ}$.

In the case of electron backscattering ($\theta_e=180^{\circ}$),
when $\tau_p=\tau_{max}$, it follows from Eqs. (\ref{tetap}) and (\ref{taum+1})
that $\cos(\theta_p)=1$ and $\theta_p=0^{\circ}$. Thus, the electron scattering
by an angle ranging from $0^{\circ}$ to $180^{\circ}$
($0^{\circ} \leqslant \theta_e \, \leqslant 180^{\circ}$)
leads to a change in the proton scattering angle from
$90^{\circ}$ to $0^{\circ}$.

\begin{figure}[h!]
\centering
\includegraphics[width=0.40\textwidth]{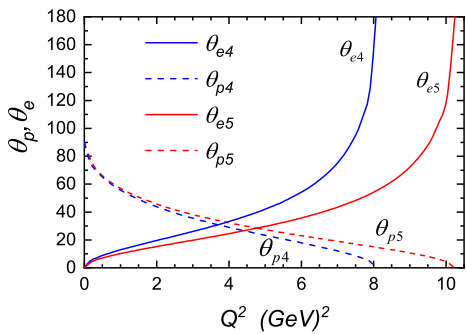}
\vspace{-3mm}
\caption{
(Color online) Electron scattering angle $\theta_e$ and proton scattering angle $\theta_p$
versus $Q^2$ at the electron beam energies used in the experiment \cite{Liyanage2020}.
The $\theta_{e4}, \theta_{p4}$, and $\theta_{e5}, \theta_{p5}$ lines are plotted for $E_1=4.725$
and $5.895$ GeV.
}
\label{Theta_ep1}
\end{figure}

The results of calculating the dependence of the electron scattering angle $\theta_e$
and proton scattering angle $\theta_p$ on the squared momentum transfer $Q^2$ to the
proton at the electron beam energies used in SANE's experiment \cite{Liyanage2020},
$E_1=4.725$ GeV and $E_1=5.895$ GeV, are plotted in Fig. \ref{Theta_ep1}. The lines with marks
$\theta_{e4}, \theta_{p4}$ and $\theta_{e5}, \theta_{p5}$ correspond to these plots.

\vspace{-6mm}

\vspace{-8mm}
\begin{table}[h!]
\centering
\caption{
Electron scattering angle $\theta_e$ and proton scattering
angle $\theta_p$ (in radians) in the kinematics of experiment \cite{Liyanage2020}
}
\label{Uglyep}
\tabcolsep=1.5mm
\footnotesize
\begin{tabular}
{| c | c | c | c | c | c | }
\hline
$E_1$ (\rm{GeV})
&  $Q^2$ (\rm{GeV}$^2$)
& $\theta_{e}\, (rad)$ & $\theta_{p}\, (rad)$
& $Q^2_{max}$ (\rm{GeV})$^2$
  \\
\hline
5.895 &2.06 & 0.27 &  0.79  & 10.247  \\
\hline
5.895 &5.66 & 0.59 & 0.43  & 10.247 \\
\hline
4.725 &2.06 & 0.35 & 0.76  & 8.066  \\
\hline
4.725 &5.66 & 0.86 & 0.35 & 8.066 \\
\hline
\end{tabular}
\end{table}

The data on the electron and proton scattering angles (in radians) at electron beam energies
$E_1=5.895$ and 4.725 GeV and $Q^2=2.06$ and 5.66 GeV$^2$ are listed in Table \ref{Uglyep},
which contains also the values of $Q^2_{max}$ (\ref{Qmax}) for the maximally possible $Q^2$
values at $E_1=5.895$ and 4.725 GeV.

{\bf Polarization of the virtual photon in the $ e p \to e p$ process.}
The $\varepsilon$ value entering the expression for the Rosenbluth
cross section,
\ba
\varepsilon=(1+2(1+\tau_p)\tan^2(\theta_e/2))^{-1}
\label{eps}
\ea
with the range of variation $0 \leqslant \varepsilon \leqslant 1$ is generally identified
in the literature with the degree of longitudinal polarization of the virtual photon.
Sometimes it is also referred to as the polarization parameter or simply the virtual photon
polarization. The physical meaning of the $\varepsilon$ value is rarely understood correctly;
in this context we should the absolutely correct words from \cite{Gakh2008}:
``Let us introduce another set of kinematical variables:
$Q^{\,2}$, and the degree of the linear polarization of the virtual
photon, $\varepsilon$''.

Expression (\ref{eps}) for $\varepsilon$ is a function of the electron
scattering angle $\theta_{e}$. An expression for $\varepsilon$, which differs
from (\ref{eps}) and makes it possible to calculate the dependences
of the quantities of interest on, e.g., $Q^2$ or the
proton scattering angle $\theta_{p}$, is given below; it was
derived using the results of \cite{GL97}:
\ba
\label{eps2}
\varepsilon^{-1}=1+\frac{(E_1-E_2)^2+2 (E_1-E_2)M} {2 E_1 E_2-(E_1-E_2)M}\,.
\ea
Here, $E_1, E_2$ are the energies of the initial and final electrons, respectively.
Note that Eqs. (\ref{E2Q}) and (\ref{E2tau}) must be used for $E_2$; they depend
explicitly on only $Q^2$; in turn, the dependence of $Q^2$ on the angles $\theta_{e}$ and
$\theta_{p}$ is determined by Eqs. (\ref{Q^2te}) and (\ref{Q^2tp}).

\begin{figure}[h!]
\centering
\includegraphics[width=0.45\textwidth]{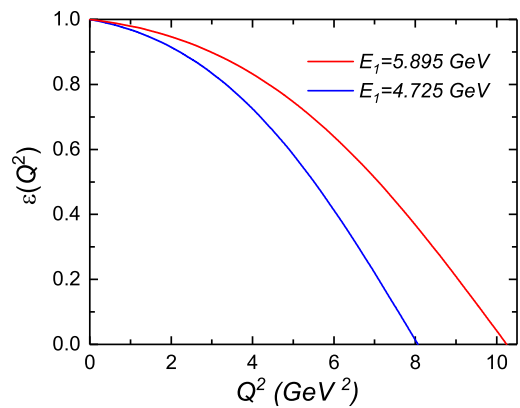}
\vspace{-2mm}
\caption{ 
(Color online) $Q^2$-dependence of the degree of linear polarization of the virtual photon,
$\varepsilon$ (\ref{eps2}), for the electron beam energies of $4.725$  and $5.895$ GeV,
used in the experiment \cite{Liyanage2020}.
}
\label{eps_01}
\end{figure}

\begin{figure*}[h!tpb]
\centering
\includegraphics[width=0.45\textwidth]{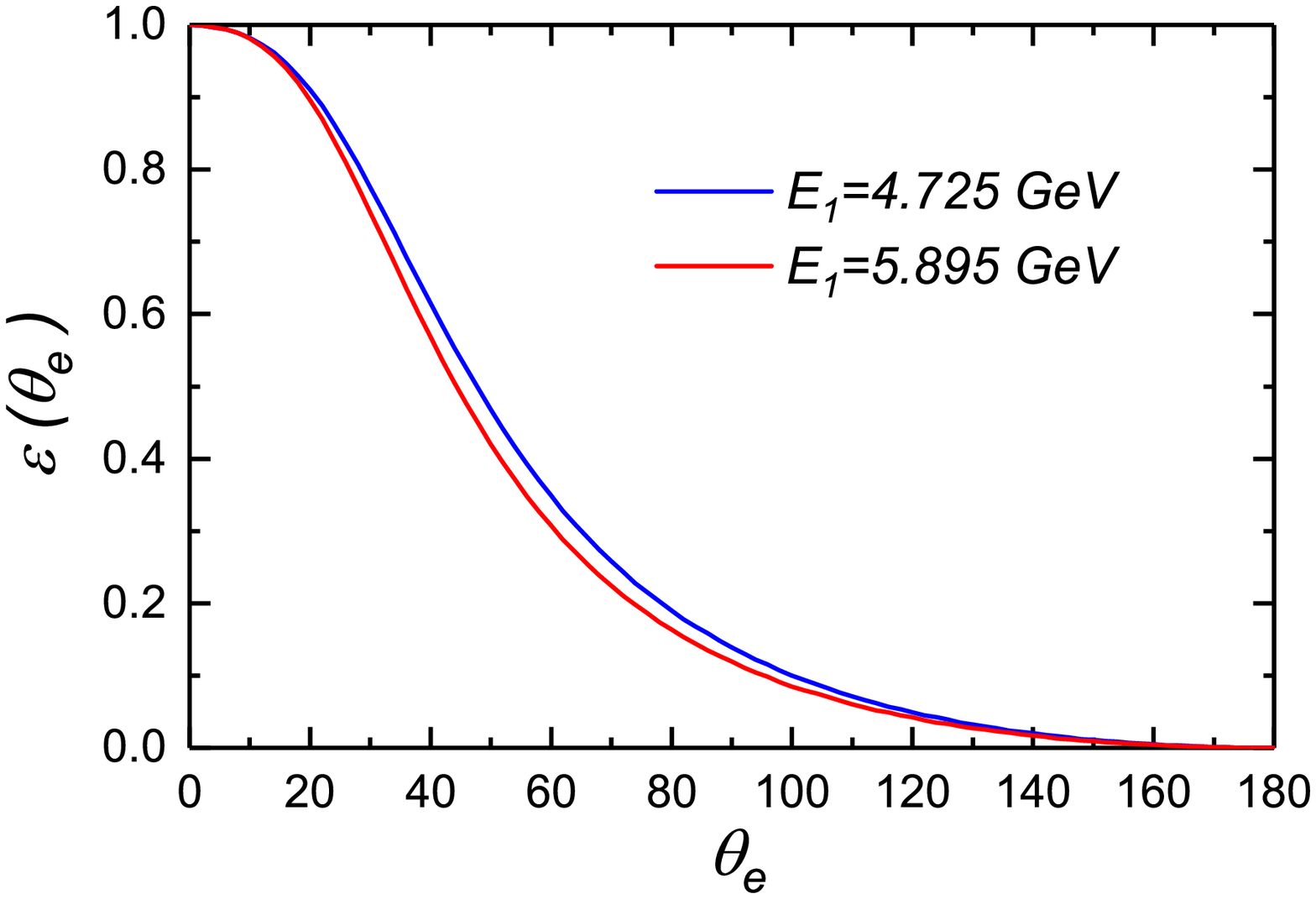}
\hspace{-3mm}
\includegraphics[width=0.45\textwidth]{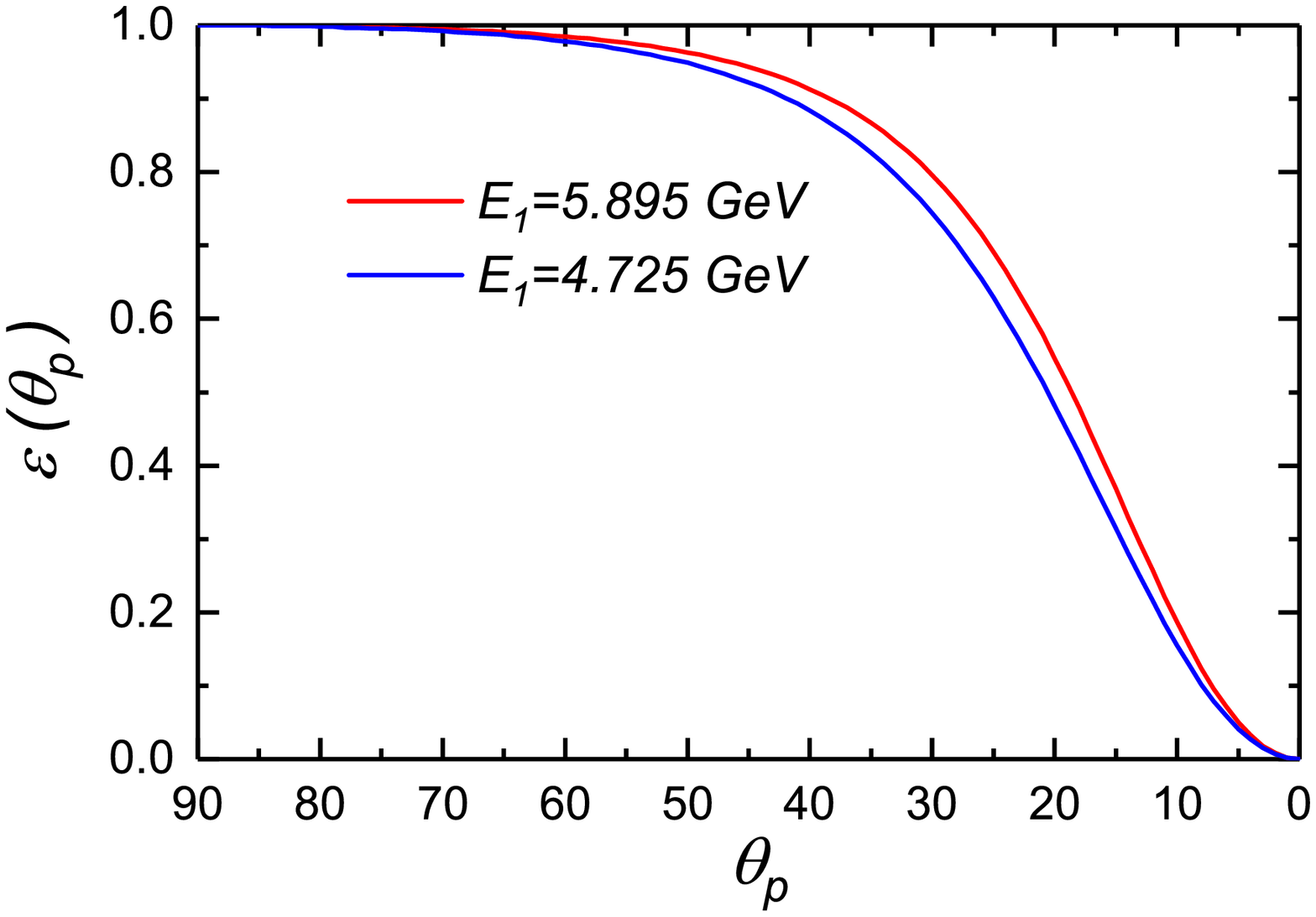}\\
\vspace{-4mm}
\caption{ 
(Color online) Degree of linear polarization of the virtual photon, $\varepsilon$,
versus the (left panel) electron scattering angle $\theta_e$ and (right panel) proton
scattering angle $\theta_p$ (in degrees) at the electron beam energies
of $5.895$ and $4.725$ GeV, used in the experiment \cite{Liyanage2020}.
}
\label{epsilon_from_theta_ep}
\end{figure*}

The dependence of the degree of linear (transverse)
polarization $\varepsilon$ (\ref{eps2}) of the virtual photon on the
squared momentum transfer to the proton, $Q^2$, at the
electron beam energies used in SANE's experiment
\cite{Liyanage2020} is presented in Fig. \ref{eps_01}.

It follows from Fig. \ref{eps_01} that $\varepsilon$ is a function of $Q^2$,
which decreases from $1$ to $0$. In the case of electron scattered forward ($\theta_e=0^\circ$),
when $Q^2=0$, $\varepsilon=1$; for a backscattered electron ($\theta_e=180^\circ$),
when $Q^2=Q^2_{max}$, $\varepsilon=0$. The $Q^2_{max}$ values for the energies $E_1=4.725$ and
$5.895$ GeV are listed in Table \ref{Uglyep}; they amount to $8.066$
and $10.247$ GeV$^2$, respectively. Specifically at these
points, the lines in Fig. \ref{eps_01} intersect the abscissa axis.

Figure \ref{epsilon_from_theta_ep} presents the dependence of the degree of
linear polarization of the virtual photon ($\varepsilon$) on the
electron and proton scattering angles ($\theta_e$ and $\theta_p$,
respectively) at the electron beam energies used in
SANE's experiment \cite{Liyanage2020}. Note that the angle $\theta_e$ in the
left panel of Fig. \ref{epsilon_from_theta_ep} changes from $0^{\circ}$ to $180^{\circ}$, while in
the right panel $\theta_p$ changes from $90^{\circ}$ to $0^{\circ}$. This order
of counting the angles $\theta_e$ and $\theta_p$ corresponds to the
range of variation $0 \leqslant Q^2 \, \leqslant Q^2_{max}$ for each plot.

The following regularities can be established based
on the plots in Fig. \ref{epsilon_from_theta_ep}: a smaller (larger) $\varepsilon$ value corresponds
to a higher electron beam energy $E_1$ for the
same angle $\theta_e$ ($\theta_p$).

{\bf Angular dependence of the polarization transfer to the proton in the $e \vec p \to e \vec p$ process.}
In the laboratory reference frame, the degree of the longitudinal polarization transfer from the initial to
the final proton in process (\ref{EPEP}) in the case of a proton target partially polarized along
the direction of motion of detected recoil proton is determined by Eq. (\ref{Asim_pt}).
Currently, the experiment aimed at measuring this parameter appears to be quite real, because such a target
with a high degree of polarization, $P_t=(70 \pm 5)$ \%, was in principle developed and even
used in SANE's experiment \cite{Liyanage2020}. For this reason it would be most expedient to perform
the proposed experiment on the facility used in \cite{Liyanage2020} at the same value $P_t=0.70$, electron
beam energies $E_1=4.725$ and 5.895 GeV, and squared momentum transfers to the proton $Q^2 = 2.06$
and 5.66 GeV$^2$. The difference between the proposed experiment and the experiment \cite{Liyanage2020}
is that the electron beam must be unpolarized, and the detected recoil proton with longitudinal
polarization must move strictly along the spin quantization axis of the proton target. This condition
stems from the requirements imposed on the spin quantization axis for both the initial and
final protons (\ref{LSO}). The procedure of measuring the degrees of longitudinal and transverse
polarizations of the final proton was developed and applied in the experiments
\cite{Jones00,Gay01,Gay02,Pun05,Puckett10,Puckett12}. To derive the $R$ ratio (\ref{Rp}) in the
proposed experiment, one must only measure the degree of longitudinal polarization of the
recoil proton, which is an advantage in comparison with the method \cite{Rekalo74} used in
\cite{Jones00,Gay01,Gay02,Pun05,Puckett10,Puckett12}.

\begin{figure*}[h!tpb]
\centering
\includegraphics[width=0.45\textwidth]{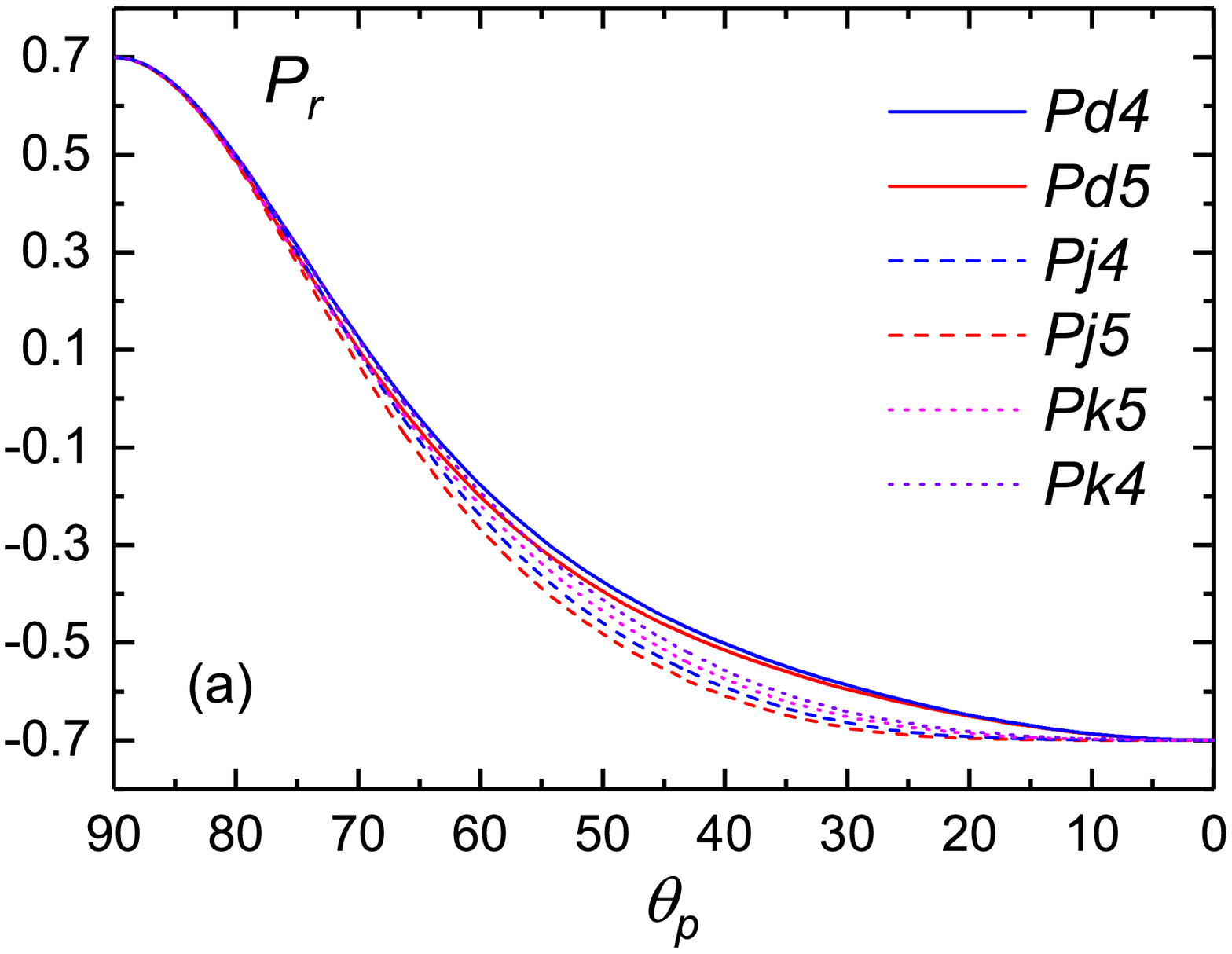}
\hspace{-3mm}
\includegraphics[width=0.45\textwidth]{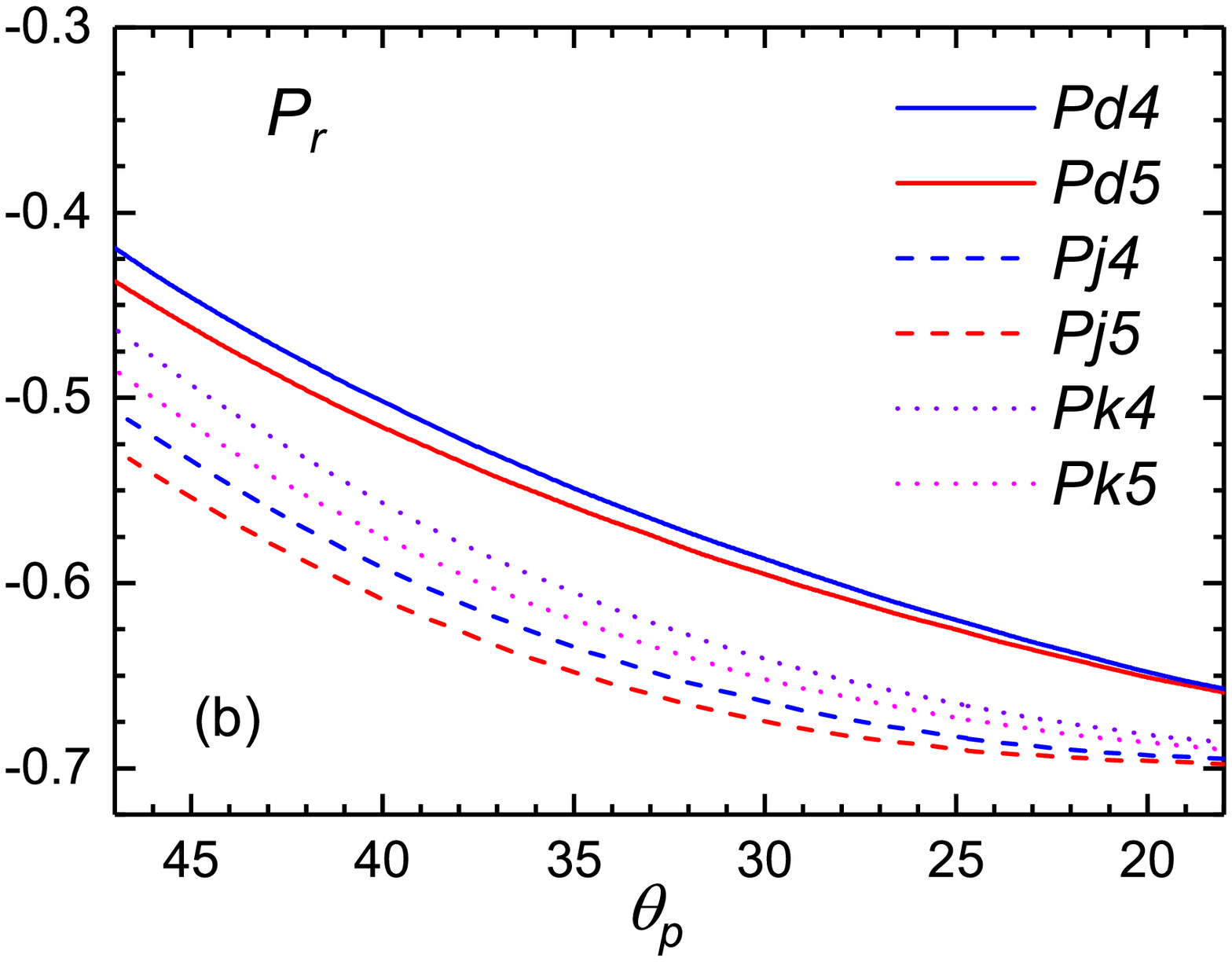}\\
\vspace{-3mm}
\caption{ 
(Color online) (a) Dependence of $P_r$ (\ref{Asim_pt}) on the proton scattering angle
$\theta_p$ for the $E_1$ and $P_t$ values used in the experiment \cite{Liyanage2020}
in the entire range of variation in the angle $\theta_p\in(90^{\circ}, 0^{\circ})$.
(b) The same dependence in the range $\theta_p\in(47^{\circ}, 18^{\circ})$, in which
$2.06 \, \rm{GeV}^2 \leqslant Q^2 \, \leqslant 5.66$ GeV$^2$.
The $Pd$, $Pj$, and $Pk$ lines correspond to the dipole dependence (\ref{Rd}),
the Qattan parameterization
(\ref{Rdj}) \cite{Qattan2015}, and the Kelly parameterization \cite{Kelly2004}.
}
\label{exp12}
\end{figure*}

The calculated dependences of the longitudinal polarization transfer to the proton,
$P_r$ (\ref{Asim_pt}), on the proton scattering angle $\theta_p$ for the electron
beam energies of $5.895$ and $4.725$ GeV and for $P_t=0.70$ are plotted in Fig. \ref{exp12}.
Figure 4a shows the dependence in the entire range of variation in the angles
$\theta_p\in(90^{\circ}, 0^{\circ})$. In Fig. 4b the range of variation
$\theta_p\in(47^{\circ}, 18^{\circ})$  corresponds to the kinematics of the experiment
\cite{Liyanage2020}, where $2.06 \; \rm{GeV^2} \leqslant Q^2 \, \leqslant 5.66$ GeV$^2$
(see Table \ref{Uglyep}). The  $Pd5$, $Pk5$, and $Pj5$ ($Pd4$, $Pk4$, and $Pj4$)
lines correspond to the electron beam energy $E_1=5.895$ ($E_1=4.725$) GeV.
In turn, the $Pd5$ and $Pd4$ lines were plotted for $R=R_d$ (\ref{Rd}) for the case
of dipole dependence; the $Pk5$ and $Pk4$ lines correspond to the Kelly parameterization
\cite{Kelly2004} ($R=R_k$); and the $Pj5$ and $Pj4$ lines were plotted for  $R=R_j$ (\ref{Rdj})
in the case of the Qattan parameterization \cite{Qattan2015}.

\vspace{4mm}

\begin{table*}[h!tpb]
\centering
\caption{  
Degree of longitudinal polarization of the recoil proton $P_r$ (\ref{Asim_pt}) at electron beam
energies $E_1=5.895$ and $4.725$ GeV and values $Q^2=2.06$ and 5.66 GeV$^2$. The values in the columns
for $Pd$, $Pj$, and $Pk$ correspond to the dipole dependence (\ref{Rd}), the Qattan
parameterization (\ref{Rdj}) \cite{Qattan2015}, and the Kelly parameterization \cite{Kelly2004}.
The corresponding electron and proton scattering angles (in degrees) are given in the columns
for $\theta_{e}$ and $\theta_{p}$}
\vspace{2mm}
\label{DeltaKelly}
\tabcolsep=2.00mm
\footnotesize
\begin{tabular}
{| c | c | c | c | c | c | c | c | c|}
\hline
$E_1$ (\rm{GeV})
&  $Q^2$ (\rm{GeV}$^2$)
&  $\theta_{e}$ (deg)
& $\theta_{p}$ (deg)
& $Pd$
& $Pj$
& $Pk$
& $\Delta_{dj}$, \%
& $\Delta_{dk}$, \% \\
\hline
5.895 & 2.06 & 15.51  &  45.23 & --0.460 & --0.552 & --0.511 & 16.6 & 9.98 \\
\hline
5.895 &5.66  & 33.57 & 24.48 & --0.628 & --0.691 & --0.675 & 9.1  & 6.96 \\
\hline
4.725 &2.06  & 19.97 & 43.27 & --0.467 & --0.556 & --0.517 & 16.1 & 9.67 \\
\hline
4.725 &5.66  & 49.50 & 19.77 & --0.649 & --0.693 & --0.682 & 6.4 & 4.84 \\
\hline
\end{tabular}
\end{table*}

It follows from the plots in Fig. \ref{exp12} that the polarization transfer to the proton depends
fairly strongly on the type of the $R$ parameterization. In the case of SFF scaling violation,
i.e., at $R=R_k$ and $R=R_j$, it essentially increases in magnitude in comparison with the case
of dipole dependence, when $R=R_d$; the following inequalities are valid for all $\theta_p$ in this
case: $|Pd5|<|Pk5|<|Pj5|$ and $|Pd4|<|Pk4|<|Pj4|$. Thus, the $Pk$ lines for the Kelly
parameterization \cite{Kelly2004} occupy an intermediate position between $Pd$ and $Pj$.

To estimate quantitatively the difference between $Pj$, $Pk$, and $Pd$, Table \ref{DeltaKelly}
was compiled, which contains the degrees of the longitudinal polarization of the final proton,
$Pj5$, $Pd5$, $Pj4$, $Pd4$,  $Pk5$, and $Pk4$, and their relative differences (in percent) $\Delta_{dj}$
and  $\Delta_{dk}$
\ba
\label{Deltadj}
\Delta_{dj}=\Big|\frac{\rm{Pd}-\rm{Pj}}{\rm{Pd}}\Big|, \;
\Delta_{jk}=\Big|\frac{\rm{Pj} - \rm{Pk}}{\rm{Pj}}\Big|\nn
\ea
at two electron beam energies (5.895 and 4.725 GeV)
and two $Q^2$ values (2.06 and 5.66 GeV$^2$).

It follows from Table \ref{DeltaKelly} that, at $Q^2=2.06$ GeV$^2$,
the relative difference between $Pj5$ and $Pd5$ is 16.6 \%;
the difference between $Pj4$ and $Pd4$ is approximately the same: 16.1 \%.
At $Q^2 = 5.66$ GeV$^2$ this difference decreases to $9.1$ and $6.4$ \%, respectively.

The difference between $\Delta_{dj}$ and $\Delta_{dk}$ in Table \ref{DeltaKelly} is small;
it varies from 2 to 6 \%. This difference can be explained by the fact that the Kelly
parameterization \cite{Kelly2004} was proposed in 2004, prior to the experiments
\cite{Puckett10,Puckett12}, whose results were taken into account in \cite{Qattan2015}
and made it possible to obtain a more accurate parameterization.

\vspace{4mm}
{\bf Conclusions.}
Proceeding from the results of JLab's polarization experiments on measuring
the ratio of the Sachs form factors in the $\vec e  p \to e \vec p$ process, using the Kelly
\cite{Kelly2004} (2004) and Qattan \cite{Qattan2015} (2015) parameterizations
for this ratio, in the kinematics of SANE's experiment \cite{Liyanage2020} (2020)
on measuring the double spin asymmetry in the $\vec e \vec p \to e p$ process, a
numerical analysis of the dependence of the longitudinal polarization transfer
to the proton in the $e \vec p \to e \vec p$ process on the proton scattering
angle was performed for the case where the initial (at rest) proton is partially
polarized along the direction of motion of the detected recoil proton.
It has been found that the polarization transfer to the proton is fairly sensitive to the
parameterization of the ratio of the Sachs form factors, which opens possibilities
for a new measurement of this ratio in the $e \vec p \to e \vec p$  process.

It follows from the calculations that the violation of the scaling of the Sachs form factors
leads to a significant increase in the magnitude of the polarization transfer to the proton,
$|P_r|$, as compared to the case of the dipole dependence; the $|P_r|$ value, obtained with
the Kelly parameterization \cite{Kelly2004}
is between the results obtained for the dipole dependence and Qattan parameterization
\cite{Qattan2015}. Obviously, the parameterization \cite{Qattan2015}, being based on a wider
set of experimental data in comparison with the Kelly parameterization, including, in particular,
the results reported in \cite{Puckett10,Puckett12}, is more accurate and objective and
leads to small differences from the results obtained with the Kelly parameterization.

\end{document}